# Investigation of the quinine sulfate dihydrate spectral properties and its effects on Cherenkov dosimetry

Emilie Jean[1,2], Marie-Ève Delage[1,2] and Luc Beaulieu[1,2]

[1] Département de physique, de génie physique et d'optique et Centre de recherche sur le cancer, Université Laval, Québec, Canada
[2] Centre de recherche du CHU de Québec – Université Laval, Québec, Canada

E-mail: emilie.jean.3@ulaval.ca

## Abstract

Recent studies have proposed that adding quinine to water while performing Cherenkov volumetric dosimetry improves the skewed percent depth dose measurement. The aim of this study was to quantify the ability of quinine to convert directional Cherenkov emission to isotropic fluorescence and evaluate its contribution to the total emitted light. Aqueous solutions of quinine were prepared with distilled water at various concentrations (0.01 to 1.2 g/L). The solutions were irradiated with photon beams at 6 and 23 MV. The dependence of the light produced as a function of sample concentration was studied using a spectrometer with a fixed integration time. Spectral measurements of the luminescent solution and the blank solution (distilled water only) were taken to deconvolve the Cherenkov and quinine contribution to the overall emission spectrum. Using a CCD camera, intensity profiles were obtained for the blank and the 1.00 g/L solutions to compare them with the dose predicted by a treatment planning system. The luminescent intensity of the samples was found to follow a logarithmic trend as a function of the quinine concentration. Based on the spectral deconvolution of the 1.00 g/L solution, 52.4% ± 0.7% and 52.7% ± 0.7% of the signal in the visible range results from the quinine emission at 6 and 23 MV, respectively. The remaining fraction of the spectrum is due to the Cherenkov light that has not been converted. The fraction of the Cherenkov emission produced between 250 nm and 380 nm in the water and that was absorbed by the fluorophore reached 24.8% and 9.4% respectively at 6 and 23 MV. X-ray stimulated fluorescence of the quinine was then proven to be the principal cause to the increased total light output compared to the water-only signal. This new information reinforces the direct correlation of the solution intensity to the dose deposition.

## 1. Introduction

Quinine sulfate dihydrate ($C_{40}H_{54}N_4O_{10}S$)(Pubchem n.d.), also named quinine, is a common organic fluorophore which is often used as a reference material for wavelength calibration (Velapoldi and Mielenz 1980). Although, mostly utilised for its pharmaceutical properties (Sullivan 2011), quinine has recently been employed to overcome a major issue encountered while performing Cherenkov volumetric dosimetry in water (Glaser *et al* 2013b), where the directionality of the Cherenkov light emission yields a skewed percent depth dose (PDD) measurement (Glaser *et al* 2013a). This effect can be corrected by taking into account the privileged emission angle (Jelley 1958, Afanasiev 2006), or as was recently attempted, by adding quinine to water following the hypothesis that it could absorb the Cherenkov light emission and convert it to isotropic fluorescent light (Glaser *et al* 2013b). Given the absorption spectrum of quinine in aqueous solution peaks between 200-250 nm, the detectable Cherenkov signal (Jelley 1958) comprised in the visible spectrum should not decrease while performing percent depth dose measurements with a CCD camera. In fact, an increase of the overall signal should be observed as the fluorophore should be able to effectively absorb undetectable ultra-violet (UV) Cherenkov light and convert it into visible fluorescence. However, being an aromatic compound, quinine inherently possesses the ability to produce fluorescence by direct excitation (i.e. intrinsic scintillator) with a photon or electron beam (Horrocks 1974). This will then contribute to the total signal collected. It is therefore questionable whether the dominant excitation source when irradiating a quinine solution results from the Cherenkov light conversion, or the luminescence stimulated directly by the ionizing radiation beam. This paper assesses the spectral properties of aqueous quinine solutions used for Cherenkov emission dosimetry. Precisely, it aims to quantify the ability of the fluorophore to convert anisotropic Cherenkov emission to fluorescent light and evaluate its contribution to the total emitted light.

## 2. Materials and methods

### 2.1. Preparation of the quinine solutions

Quinine sulfate dihydrate is a crystalline white powder which is soluble in water at concentrations up to 1.23 g/L (Fisher Scientific 2013). In this paper, we have explored the effect on Cherenkov dosimetry of various concentrations, between 0.01 g/L and 1.2 g/L, of quinine in aqueous solutions. Being sensitive to light and pH, solutions were prepared with distilled water immediately before being used to prevent potential degradation of the molecule. The quinine, which was weighed beforehand, was successively added to a fixed volume of distilled water so the concentration of the solution could be gradually increased. With this procedure, the same container was used for all measurements, ensuring uniformity of the solution signal and the Cherenkov production in the tank itself. The powder volume necessary to achieve a maximum concentration of 1.23 g/L in 600 ml of distilled water was only 0.92 ml due to its density of 0.8 g/ml (Fisher Scientific 2013). This volume was negligible compared to the total volume of the water used for the solution. Moreover, quinine does not possess any high atomic number elements, so the prepared solutions should posses similar properties (refractive index, attenuation coefficient, etc.) to water.

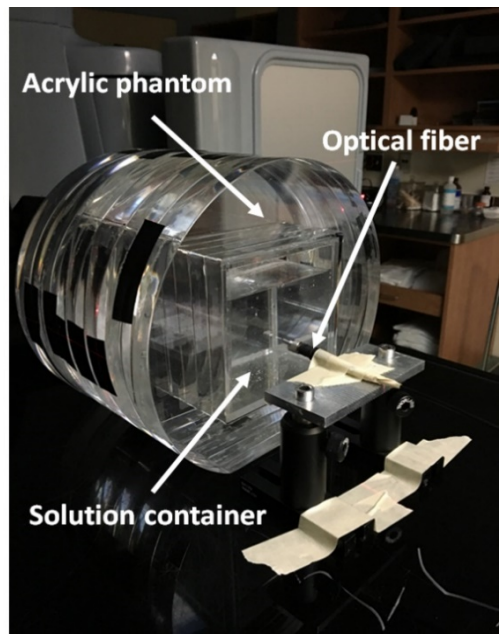

*Figure 1 : Acrylic cylindrical phantom with the central 10x10x10 $cm^3$ sensitive volume and the optical fiber affixed to the container's side for spectral measurments.*

### 2.2. Spectral properties measurement set-up

The quinine solutions were irradiated with a linear accelerator photon beam (Clinac iX, Varian Medical Systems, Palo Alto, USA) at 6 MV and 23 MV with a field size of 40 x 40 $cm^2$ and a constant dose rate of 600 MU/min. The container used for the irradiation was an acrylic cube of 10 cm length. Due to the wall thickness, it has a total capacity of 753.6 ml but was only filled with 600 ml of solution. The container was placed in a cylindrical acrylic phantom (Goulet *et al* 2014) and centered on the treatment isocenter as illustrated in Fig. 1. An optical PMMA fiber (Eska GH-4001, Industrial Fiber Optics, Tempe, USA) of 15 m long with a diameter of 1 mm and a 0.5 numerical aperture was connected to a UV and shortwave sensitive spectrometer (QE65Pro, Ocean Optics, Dunedin, FL). The spectrometer was equipped with a 50μm slit and the grating H3 (600 lines/mm blazed at 500 nm) for all measurements and was placed outside the treatment room. The fiber was affixed directly on the container's side for spectral measurements of all the solution concentrations. Then, the container filled with 600 mL of distilled water was irradiated under the same conditions mentioned before to quantify the Cherenkov emission produced in the solvent (water). For all measurements, black blankets covered the set-up to avoid signal contamination by ambient light. Even with this precaution, a background spectrum measurement was made after each irradiation measurement. A spectrum was also obtained at both beam energies with the empty container in place in the set-up. This way, the Cherenkov light produced by the optical fiber, the container and the acrylic phantom placed in the radiation beam path was collected. The background and Cherenkov spectral contribution from the set-up components were subtracted from all the spectra.

*2.2.1. Dependencies of the spectral emission intensity*

Since the fluorophore is intended to be used for dosimetry, the influence of the dose, fluorophore concentration and beam energy on the spectral emission intensity has been quantitatively studied by varying successively these specific parameters. The intensity of the output signal as a function of the dose was studied by irradiating the 1.00 g/L solution with exposure times of 5, 10 and 15 seconds. Then, the effect of the fluorophore concentration on the total light emitted was investigated using nine different concentrations from 0.01 g/L to 1.2 g/L with a 10-second irradiation. The intensity of the total light emitted was obtained by integrating the spectrum counts between 380 nm and 1000 nm, which corresponds to the wavelength range where signal could be measured by the spectrometer coupled with the PMMA fiber. It also matches the CDD sensible wavelength range that was used to measure PDD which does not cover the quinine absorption spectrum region. Each measurement has been performed at both 6 MV and 23 MV to determine if the energy employed to irradiate the sample had an impact on the result.

*2.2.2. Quinine and Cherenkov light contributions*

The contribution of the quinine and Cherenkov light to the overall spectrum were first determined in the visible range only. The deconvolution of the solution spectra was obtained by a least square fitting of a linear combination of the Cherenkov and quinine spectra. Spectral measurements of the quinine emission in absence of Cherenkov light were performed using the same container and a narrow-band UV light peaked at 360 nm as the excitation source. It was found that aqueous quinine solution does not emit any fluorescence photons between 630 nm and 710 nm (spectrum available in the Supplementary Material section) as opposed to Cherenkov emission spectrum which is continuous. Measurements made with the water only were used for the Cherenkov spectrum input. The spectral deconvolution of the total emitted light from all the solutions allowed us to quantify the contribution of both Cherenkov light and quinine fluorescent light. To achieve this, the quinine fluorescence, Cherenkov and solution emission spectrum counts were integrated from 380 nm to 1000 nm.

*2.3. Cherenkov absorption*

To determine the ability of the fluorophore to convert anisotropic Cherenkov light to isotropic fluorescent light, measurements were performed in the UV range using a UV-VIS sensitive CCD camera (PiMax4 1024f, Princeton Instruments, Thousand Oaks, USA) coupled with a spectrometer (Isoplane 160, Princteon Instruments, Thousand Oaks, USA). The spectrometer was equipped with a 50μm slit and a standard grating (300 gr/mm blazed at 500 nm) and was placed inside the treatment room for all measurements. Lead shielding was surrounding the camera to minimize the noise. The solarization resistant fiber (FG400AEA, Thorlabs, Newton, USA) of 2 m long with a diameter of 0.4 mm and a 0.22 numerical aperture was placed directly on the solution surface forming a 45 degrees angle with the latter and facing the gantry that was oriented at 90 degrees to maximize the collected Cherenkov signal. For all measurements, black blankets covered the set-up to avoid signal contamination by ambient light. Irradiations at 6 MV and 23 MV using a 15 x 15 $cm^2$ field size were performed with a large acrylic tank containing 2.5 L of 1.00 g/L quinine solution. The side of the tank facing the gantry was located at a source surface distance of 98.5 cm and all inner walls were covered with a thin black opaque film. Measurements with the same volume of water were performed under the same irradiation conditions. A spectrum was also obtained at both beam energies with the empty container in place in the set-up and was subtracted from all further spectra. As quinine does not emit in the UV range, all spectra measured from 250 nm to 380 nm should be composed of Cherenkov emission only. Therefore, the water spectrum was directly subtracted from the solution spectrum and negative intensity counts below 380 nm in the remaining spectrum were considered as Cherenkov light absorbed by the fluorophore.

*2.4. Comparison with commercial scintillators*

Vials filled with 20 mL of quinine solutions with various concentrations (0.01, 0.10 and 1.00 g/L) and two commercial liquid scintillators, Utlima Gold (Perkin Elmer, Waltham, USA) and Cytoscint (MP Biochemicals, Santa Ana, USA), were exposed at 6 MV. Solid water was surrounding the vial to ensure electronic equilibrium. The same dose was given to each scintillator in order to compare their absolute light emission intensity. A polychromatic cooled CCD camera (Alta U2000, Apogee, Roseville, USA) was used to collect the light output of each scintillator using 10-second signal integrations in a 40 cm field size at 600 MU/min. The camera was coupled with a variable focal length objective (f50mm/Aperture 0.95-16, JLM optical, Rochester, USA) with the aperture set at 16 and focus on infinite. All images acquired were kept in their original raw format and the three channels (Red, Green, Blue) were taken for analysis. As the three fluorophore emission spectra are similarly peaking in the blue region, no corrections were made to consider the variation of the CCD response as a function of the channels. Using a mirror placed at 45˚, the camera could be perpendicular to the vial at 18 cm as shown in Fig. 2. Because of the proximity of the LINAC beam, lead shielding was surrounding the camera to prevent any damage to the chip from scattered radiation. The output signal was obtained by integrating the CCD pixel intensity values in an equal sized region of interest located at the centre of the solution vial.

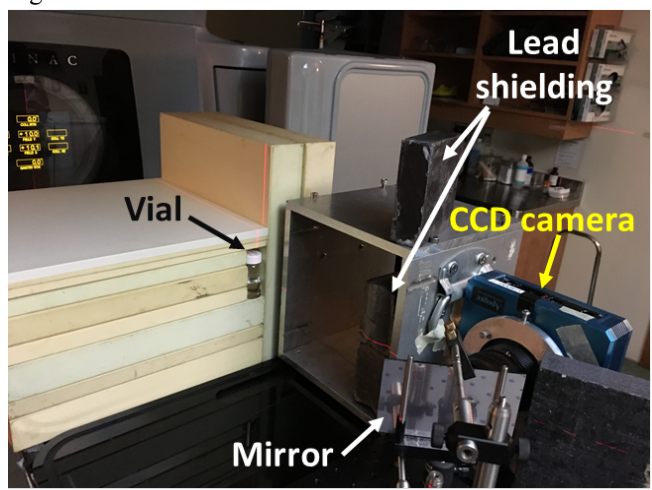

*Figure 2 : Set-up used for the comparison with commercial liquid scintillators showing the position of the CCD camera.*

### 2.5. PPDs and dose profiles

Since the main role of the quinine is to provide a correction to the skewed percent depth dose measurement (Glaser et al 2013a), dose profile and PDD measurements were obtained by irradiating a larger acrylic tank (15 x 15 x 20 cm3) filled with a volume of 17 cm height of distilled water. A field size of 5 x 5 cm2 and a constant dose rate of 600 MU/min were used at 6 MV and 23 MV. The tank was placed to obtain a source-surface distance of 100 cm. The same CCD camera used previously was placed perpendicularly to the beam's path at 85 cm from the extremity of the tank to collect the light outputs as shown in Fig. 3. Respectively ten measurements of the Cherenkov emission and the background were performed using 20-second integration time in order to apply a temporal median filtering, as the noise collected had a noticeable contribution to the overall signal. This noise reduction technique has proven in previous publication to provide great results (Archambault et al 2008). Besides, the camera was at a large distance from the field. It was then decided not to use any shielding to protect the CCD chip from scattered radiation in this set-up configuration. The same procedure was repeated with the 1.00 g/L quinine solution. To minimize light reflection inside the tank, the inner walls, except the one facing the CCD, were covered with a thin black opaque film.

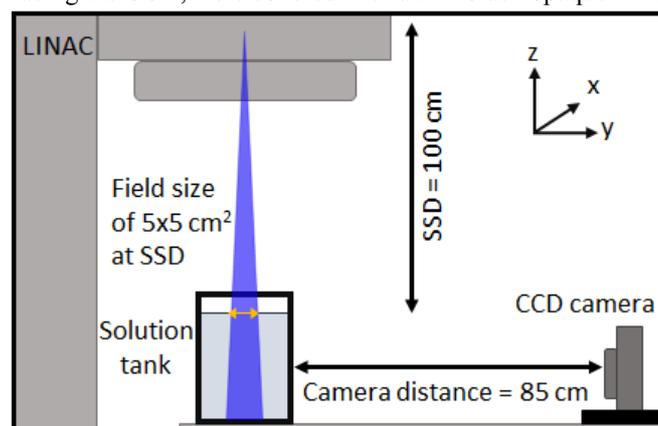

*Figure 3 : Schematic of the set-up used for the PDD and profile measurements showing the definition of the coordinates chosen for the TPS dose summation.*

After applying the median temporal filter and subtracting the background pixel value from raw images, outliers above 2 standard deviations were replaced by the surrounding mean pixel values. The pixel intensity values were extracted along the central axis, Z, of the beam to compare the PDDs, while profiles in the X direction were taken at the measured maximum dose depth (1.20 and 3.05 cm at 6 MV and 23 MV respectively). The set-up was then scanned using a computed tomography scanner (Somatom Definition, Siemens, Erlangen, Germany). The images were exported into the treatment planning system Pinnacle3 v9.8 (Phillips, Amsterdam, Netherlands) in order to predict the dose distribution at each geometrical point. As the objective parameters were set to obtain a depth of field covering the tank thickness, the light collected by the CCD is equal to the sum of the optical photons produced over the tank thickness. Thus, the 2D dose grids produced in the XZ planes of the TPS were consequently summed over the Y axis length. To facilitate the comparison, all PDDs and dose profiles were normalized to their respective maxima.

## 3. Results and discussion

### 3.1. Dependencies of the spectral emission intensity

Spectral intensities, from which the background signal and Cherenkov emission from the set-up were subtracted, were integrated over all wavelengths. This calculation was performed for the solution concentrations mentioned in the Material and Methods section. Both 6 MV and 23 MV data were analyzed. The total light intensity collected at each energy was found to be linear with exposure time, hence with the dose deposited in the solution as shown in Fig. 4 for the 6 MV beam. The data plotted for the 23 MV beam in Fig. S3 can be found in the supplementary material (SM). This proportionality between dose and intensity is a commonly sought after dosimetric property as it facilitates the ease at which dose deposition analysis can be performed.

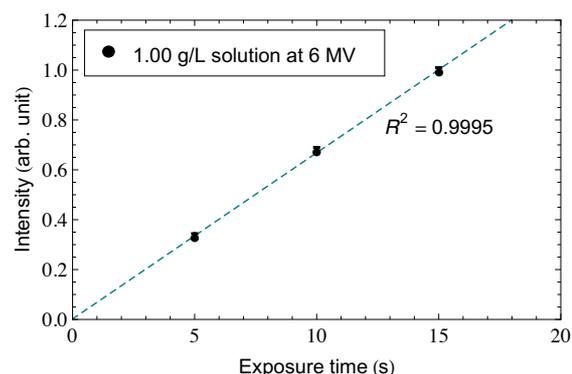

*Figure 4 : Integrated spectral intensities measured for the 1.00 g/L concentration as a function of the exposure time at 6 MV.*

Furthermore, the total intensity collected for a 10-second exposure time was found to follow a logarithmic trend as a function of fluorophore concentration at both energies, as shown in Fig. 5 for 6 MV. The quinine light emission efficiency stays around its maximal value way before its solubility limit (1.23 g/L). Therefore, there is no significant gain in intensity for concentrations higher than about 0.60 g/L. The same observation can be made with the 23 MV data plotted in Fig. S3 available in the SM. Moreover, the upper solubility limit no longer constitutes the main constraint when aiming to increase the light production.

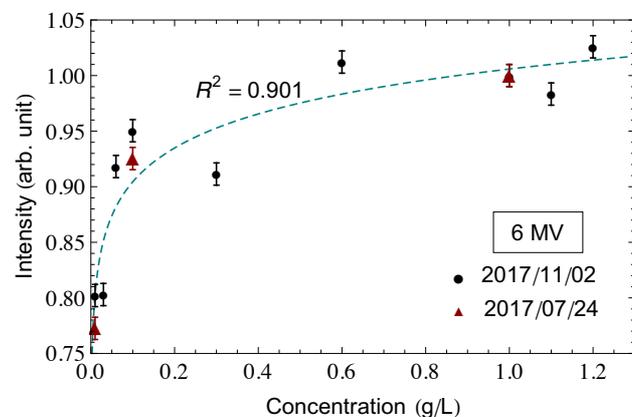

*Figure 5 : Integrated spectral intensities measured for a fixed time irradiation at 6 MV as a function of various concentration of the quinine solution.*

Finally, the ratio of the total light collected at 6 MV and 23 MV as a function of concentration, as displayed in Fig. 6, was found to be constant. As predicted, increasing the energy of the incident photon beam results in an increase in the total amount of light collected. The correlation between these two parameters was found to be independent of the concentration of the quinine solution.

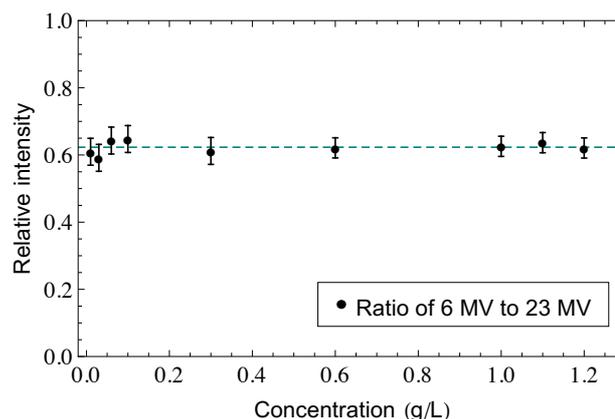

*Figure 6 : Ratio of 6 MV to 23 MV integrated spectral intensities measured for a fixed time irradiation as a function of the concentration of the quinine solution*

### 3.2. Quinine and Cherenkov light contributions

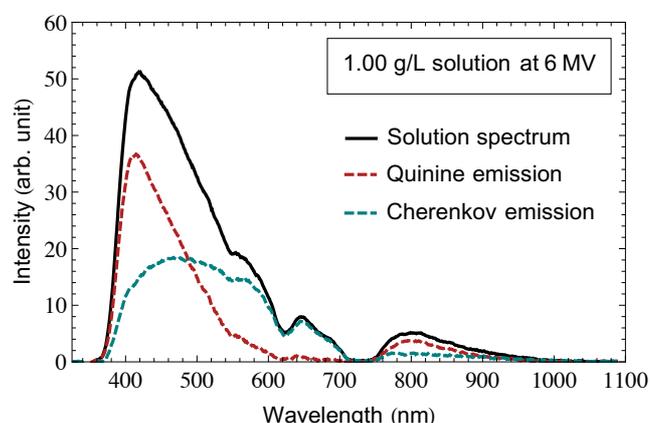

*Figure 7 : Deconvolution of the 1.00 g/L quinine solution spectrum at 6 MV for a 10-second irradiation.*

The deconvolution of the total emitted light spectrum of the 1.00 g/L solution showed that 52.4% ± 0.7% and 52.7% ± 0.7% of the signal results from quinine emission at 6 MV and 23 MV, respectively. The remaining fraction of the solution spectrum is attributable to the Cherenkov light emission produced in water that was not absorbed by the fluorophore. The contribution of the Cherenkov emission that is still present in the solution spectrum can be observed in Fig. 7 for the 6 MV energy beam and in Fig. S4 of the SM for the 23 MV beam. The clear cut off that can be seen at 380 nm is due to the resolution of the spectrometer, the grating and the absorption spectrum of the PMMA fiber, which also causes the bulges in the spectrum. The fiber absorption was not deconvolved since a relative analysis of the solution emission components was performed and the exact same fiber was used for all measurements. The deconvolution of the fiber absorption would provide spectra with different shapes and a higher intensity. Although, the relative intensity between Cherenkov and quinine emissions would remain the same as they are similarly affected by the absorption spectrum.

Fig. 8 illustrates the percentage of the Cherenkov light measured in the blank solution that was not present in the quinine solution spectrum for all concentrations tested at 6 MV. The plotted data for the 23 MV beam are presented in Fig. S5 of the SM. As quinine absorption spectrum does not extend in the wavelength range of measurements, it was therefore attributed to an increase of the quinine solution opacity. This was confirmed with a transmittance measurement made with a spectrophotometer (Cary 50, Varian, Palo Alto, USA). Quinine has shown the ability to attenuate equally on all wavelengths up to 7% of the Cherenkov produced in the water. The Cherenkov emission remaining in the 1.00 g/L solution then represents 93% of the blank solution signal, but only 47.6 % of the quinine solution signal (see Fig. 7) because the quinine has increased the total light output compared to water only.

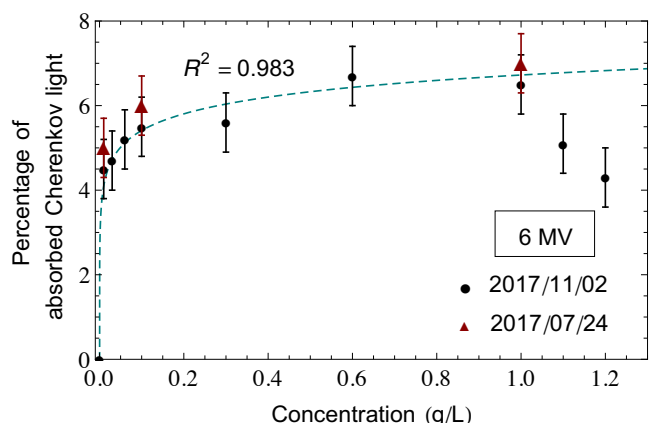

*Figure 8 : Fraction of the Cherenkov emission light produced in the blank solution attenuate by the fluorophore measured for fixed irradiation time at 6 MV as a function of various concentrations of the quinine solution.*

The fraction of Cherenkov light attenuated by the opacification of the quinine solution was found to follow an increasing logarithmic trend as a function of quinine concentration at both energies. However, the trend fails when the quinine solution approaches its solubility limit, resulting in a significant reduction of the Cherenkov light attenuation. This observation cannot be attributable to an ineffective dissolution of the quinine in water. A poor dissolution would lead to an accumulation of quinine powder at the bottom of the container, resulting in a lower concentration in the middle where the fiber was collecting the light output. As the concentration was increased successively, a poor dissolution near the solubility limit would have provided similar results to those obtained for the 1.00 g/L solution since this concentration was already reached. In fact, the attenuation would not have decreased, but rather have stayed around a fixed value. Furthermore, the same results are obtained while performing measurements at different energies. Also, measurements taken at a later date display this same behaviour, meaning the results are reproducible.

### 3.3. Cherenkov absorption

The subtraction of the water spectrum to the 1.00 g/L solution spectrum obtained at 6 MV with the UV measurements showed that an absorption of the Cherenkov light occurred below 380 nm as illustrated in Fig. 9. The same observation can be made with the 23 MV beam data presented in Fig. S6 of the SM. The negative portion of the difference spectrum corresponds to the Cherenkov light that has effectively been absorbed by the fluorophore and converted to isotropic fluorescent light. By integrating the intensity counts under 380 nm for all spectra, it was determined that the absorption represents 24.8% of the water spectrum at 6MV. However, measurements made with the 23 MV energy beam showed that the absorption only reached 9.4% of the water spectrum. Taking into account the quantum yield of quinine (0.55), 40.3% and 27.1% of the detectable quinine fluorescence between 380 nm and 450 nm results from the Cherenkov light conversion at 6 MV and 23 MV, respectively. The rest being fluorescence stimulated directly by the ionizing radiation beam, which makes the latter the major contributor to the quinine light emission as expected. Indeed, the energies implied in the numerous interactions with charged particles in the surrounding medium are greater than those carried by the few Cherenkov photons emitted, which ease the energy transfer. It is important to mention that Cherenkov and quinine emission spectra extend up to 1000 nm and are therefore not entirely covered by the UV measurements. Consequently, the percentage of absorbed Cherenkov light and fluorescent emission attributable to the Cherenkov conversion would be much smaller if the entire spectra were used for the ratios.

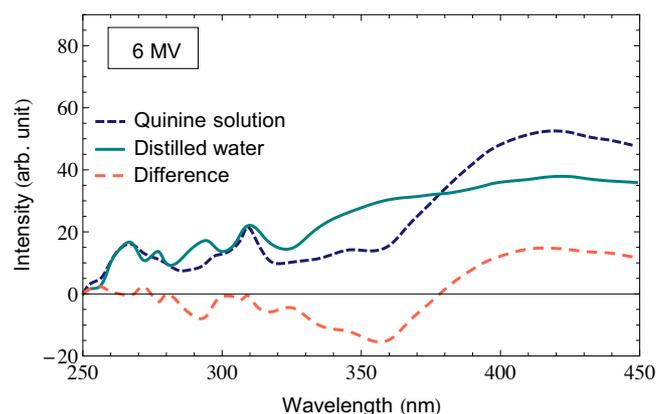

*Figure 9 : Spectra obtained for the blank solution and the 1.00 g/L quinine solution at 6 MV for a 10-second irradiation. The orange dashed line illustrates the difference between the latters. The negative intensity counts are attributable to Cherenkov light absorption while the quinine emission spectrum is clearly located above 380 nm.*

*3.4. Comparison with commercial scintillators*

The CCD measurements made with two commercial liquid scintillators have demonstrated the weak propensity of the quinine to emit luminescence. Under the same irradiation conditions, the commercial scintillator Ultima Gold (UG) produced the highest light intensity of all scintillators tested. It was therefore used as a reference to form a basis of comparison with the other scintillators. We have found that the CytoScint emitted 92.6% ± 0.1% of the UG intensity, while quinine emitted only 1.85% ± 0.07%, 1.96% ± 0.02% and 2.09% ± 0.01% of the UG intensity for concentrations of 0.01 g/L, 0.10 g/L and 1.00 g/L respectively. Since the intensity stays around a maximal value before its solubility limit (see Fig. 5), measured light intensity of the quinine solution could only be increased by optimising light collection.

*3.5. PDDs and dose profiles*

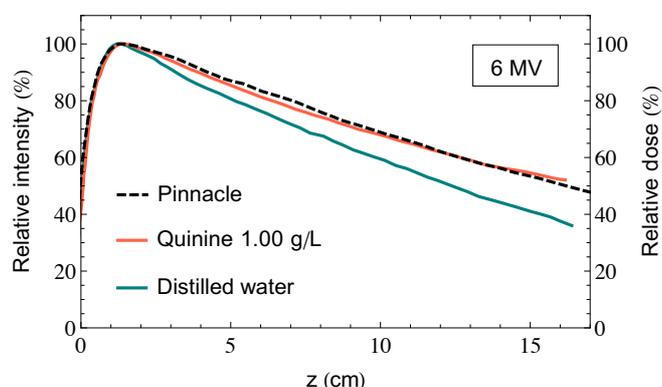

*Figure 10 : Sum of the PDDs predicted by the TPS over the tank thickness compared to the intensity fitting curves obtained along the Z axis of the CCD measurements at 6 MV for the distilled water only and the 1.00 g/L quinine solution.*

The pixel intensity values on the central axis of the beam's path obtained from the blank solution and the 1.00 g/L solution for 6 MV beam are displayed in Fig. 10. Also, the PDD summation over the tank thickness (Y axis) as predicted by the TPS are shown on the same graphic. A difference between the collected intensities and the predicted dose can be observed for both water and the quinine solution. However, measurements made with distilled water widely misrepresent the dose deposition by underestimating it systematically. The Cherenkov light emission leads to a discrepancy of the predicted dose that reaches 13% at 6 MV and 23 MV as it can be seen in Fig. S7 of the SM. Despite that quinine solution does not perfectly reproduce the expected results, it has effectively improved the PDD measurements and lowered the relative error to within 3% after the maximum dose depth for both energies employed. Even though new observations about the Cherenkov absorption process has been made, results obtained for the dose measurements are consistent with those found in the literature (Glaser *et al* 2013b).

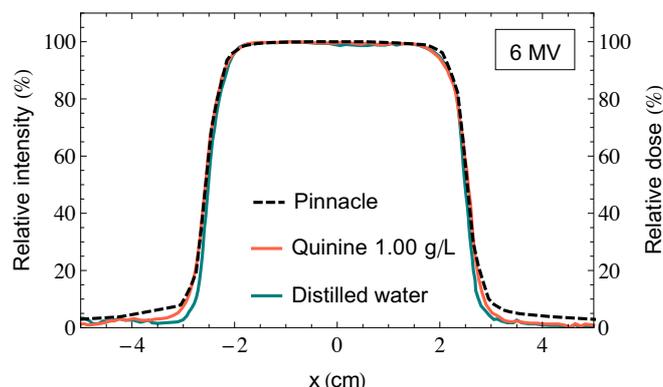

*Figure 11 : Sum of the profiles predicted by the TPS over the tank thickness compared to the intensity fitting curves obtained in the X direction at the maximum dose depth for a 6 MV energy beam.*

Similarly, the light intensity taken in the cross-profile direction at the maximum dose depth revealed a slight deviation between the TPS predicted dose and the solution light intensities at 6 MV. According to Fig. 11, it appears that the quinine solution intensity reflects the dose deposition to within 4 % in the full width half maximum (FWHM) and penumbra regions. On the other hand, the Cherenkov-only measurement displays a difference reaching 12.5%. The dose in the tail region is misrepresented by both the Cherenkov and the quinine solution intensities. Nevertheless, the difference is kept within 5% for the latter in all the profile regions. The results obtained at 23 MV provided in Fig. S8 in the SM are comparable, aside from the tail regions where the intensity of the solution reflects the dose deposition to within 3% and does not exceed 4% in the penumbra and FWHM regions. Also, the Cherenkov intensity is closer to the TPS prediction than it was at 6 MV, with a maximum deviation of 7%. In both cases, the PDD and profile improvements can be attributable to the isotropic fluorescent light contribution since it represents more than half of the total light output collected by the CCD.

These results provide an interesting explanation on how quinine improves the percent depth dose measurement quality. It increases considerably the total light output rather than acting as a Cherenkov energy transfer intermediate. While using only water, the anisotropic Cherenkov provides a skewed PDD as predicted, due to the privileged direction of the emission. When adding quinine to water, since more than half of the signal comes from the isotropic fluorescence, the proportional correlation between the dose deposition and the light intensity collected along the beam path is reinforced. Moreover, it is clear at this point that adding quinine to water while performing Cherenkov volumetric dosimetry makes X-ray stimulated fluorescence the major contributor to the dose measurement. In such a scenario, this technique becomes volumetric scintillation dosimetry.

## 4. Conclusions

The aim of this study was to quantify the ability of quinine to convert anisotropic Cherenkov emission to fluorescent light, and evaluate its contribution to the total emitted light. Quinine has shown the ability to convert less than 24.8% and 9.4 % of the Cherenkov emission produced by the water only in the UV range for the 6 MV and 23 MV beam, respectively. It represents 40.3% and 27.1% of the detectable quinine fluorescence by the UV measurements respectively at 6 MV and 23 MV. As quinine and Cherenkov emission spectrum extend up to 1000 nm, these percentages would significantly decrease if the intensity emitted over the entire spectra was used for the ratios. Furthermore, based on the spectral deconvolution of the 1.00 g/L solution in the visible range that matches the CCD sensible wavelength, 52.4% ± 0.7% and 52.7% ± 0.7% of the signal results from the quinine emission at 6 MV and 23 MV, respectively. Although quinine does not absorb in the visible range, an attenuation of the Cherenkov light produced in the water was observed due to the opacification of the quinine solution, increasing logarithmically with the concentration. Since only a slight fraction of the Cherenkov emission is converted by the fluorophore, the evidence suggests that the privileged light emission process of the quinine results from interactions with the linear accelerator photon beam. Consequently, the PDDs improvement can be linked to the fact that over half of the signal is fluorescent light, which is known to be isotropic. These results complete prior works and facilitate the comprehension of the mechanisms behind the PDD and profile quality improvement when quinine is added to water while performing Cherenkov volumetric dosimetry.

## Aknowledgments

This work was financed by the Natural Sciences and Engineering Research Council of Canada (NSERC) Discovery grants #435510-2013

...

## Supplementary material

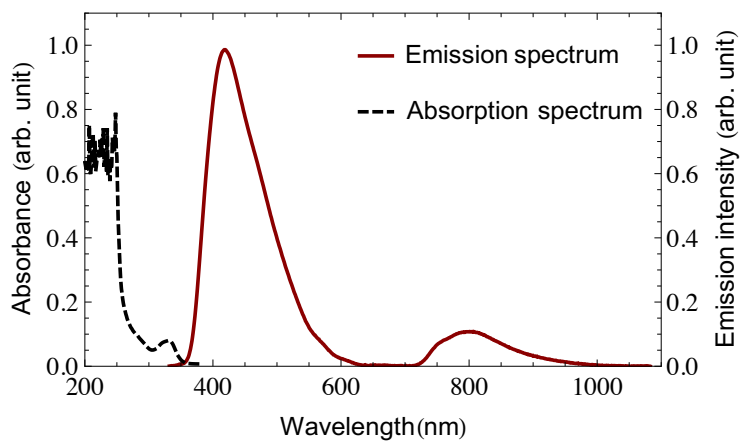

*Figure S1 : Absorption and emission spectrum of the aqueous quinine sulfate dihydrate solution. The emission spectrum was obtained using a narrow-band UV light peaked at 360 nm as the excitation light source.*

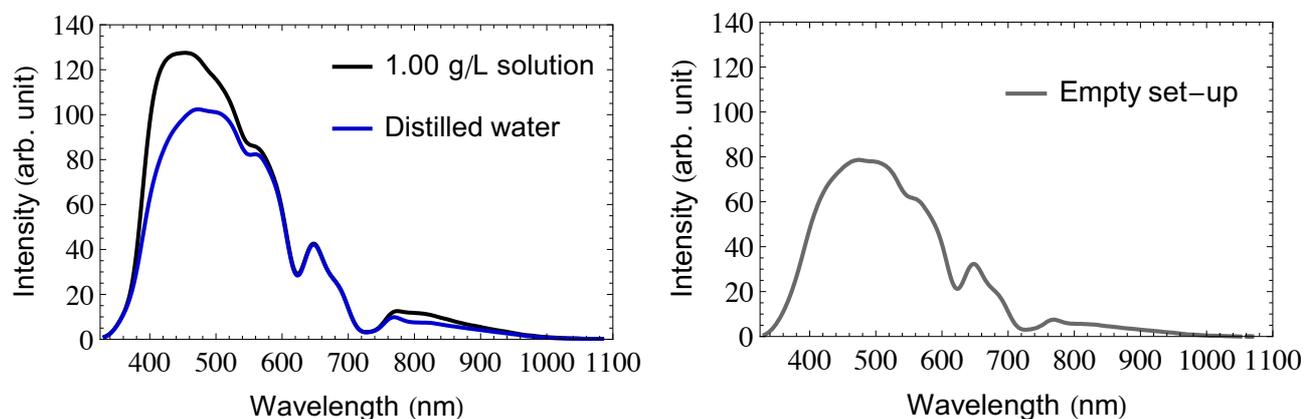

*Figure S2 : On the left, 1.00 g/L solution and distilled water raw spectra obtained with the spectrometer intensity counts and, on the right, the background Cherenkov emission raw spectrum from the tank, the fiber and the phantom for a 10-second irradiation time at 6 MV. The latter background is removed from all other measurements.*

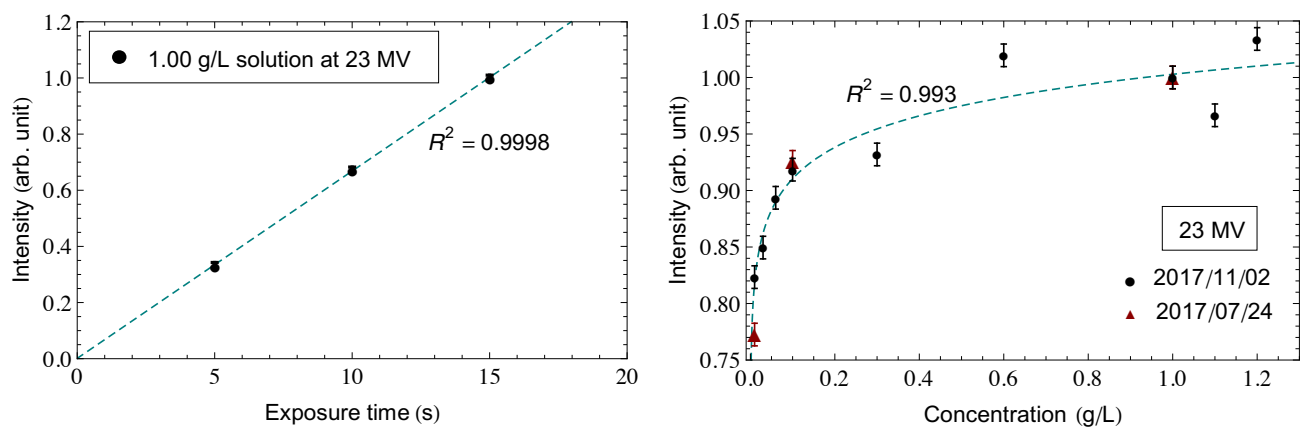

*Figure S3 : On the left, integrated spectral intensities measured for the 1.00 g/L concentration as a function of the exposure time at 23 MV. On the right, integrated spectral intensities measured for a fixed time irradiation at 23 MV as a function of various concentration of the quinine solution.*

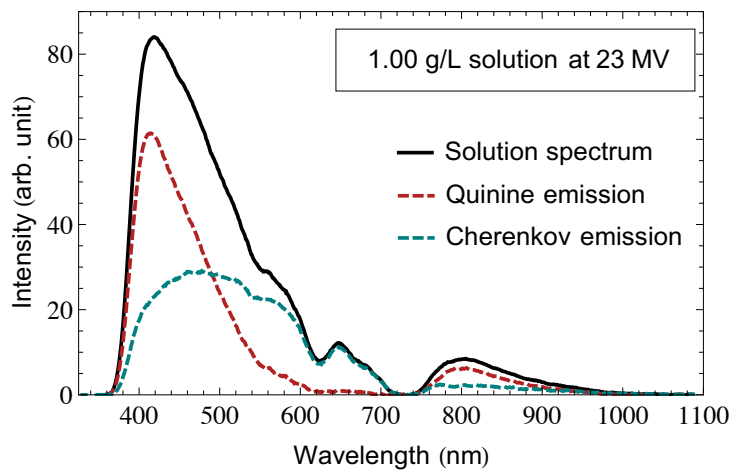

*Figure S4 : Deconvolution of the 1.00 g/L quinine solution spectrum at 23 MV.*

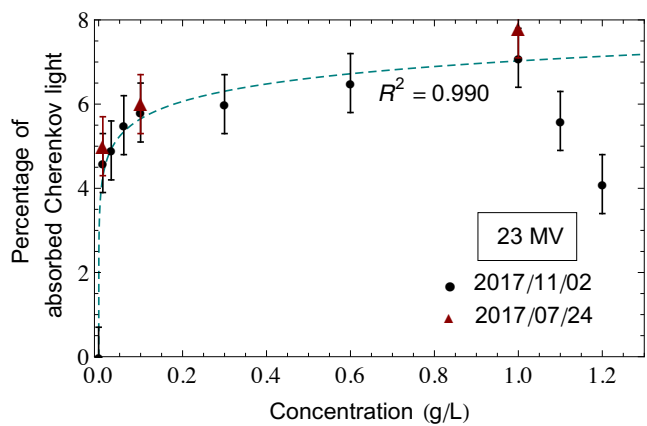

*Figure S5 : fraction of the Cherenkov emission light produced in the blank solution attenuated by the fluorophore measured for fixed irradiation time at 23 MV as a function of various concentration of the quinine solution.*

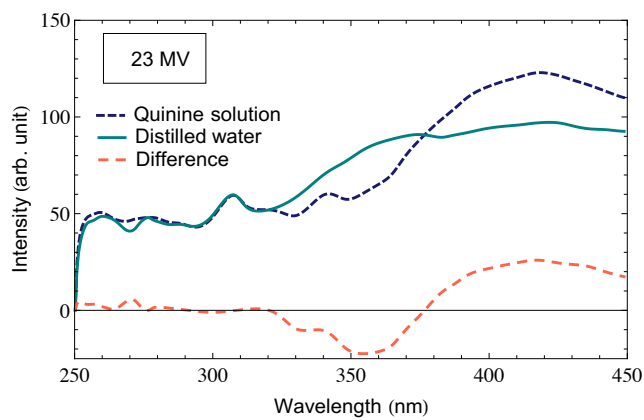

*Figure S6 : Spectra obtained for the blank solution and the 1.00 g/L quinine solution at 23 MV for a 10-second irradiation.The orange dashed line illustrates the difference between the latters.*

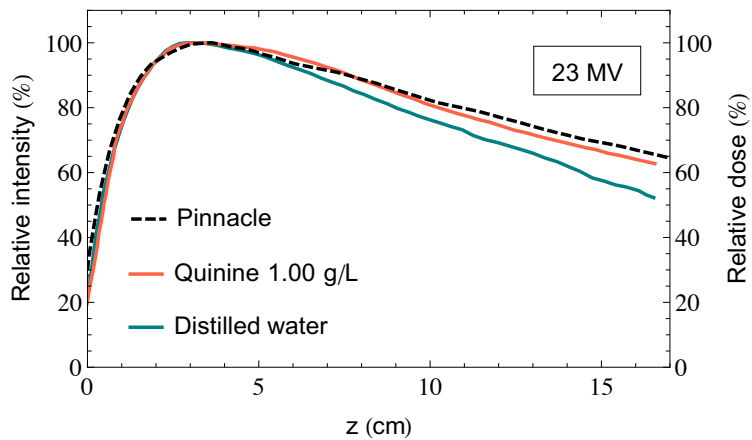

*Figure S7 : Sum of the PDDs predicted by the TPS over the tank thickness compared to the intensity fitting curves obtained along the Z axis of the CCD measurements at 23 MV for the distilled water only and the 1.00 g/L quinine solution.*

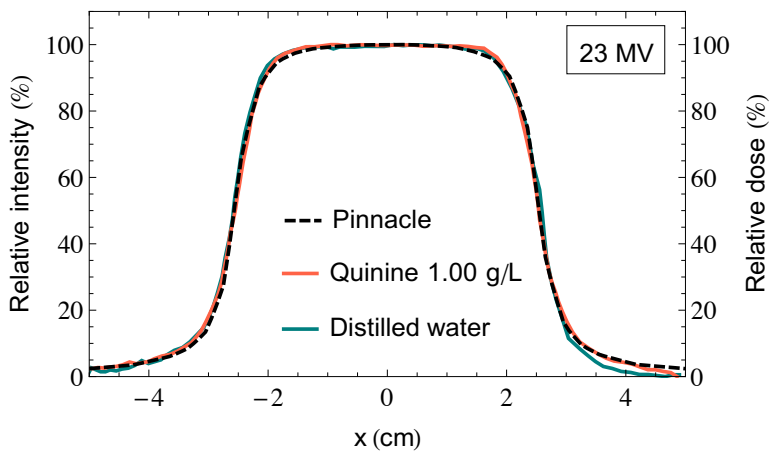

*Figure S8 : Sum of the profiles predicted by the TPS over the tank thickness compared to the intensity fitting curves obtained in the X direction at the maximum dose depth for a 23 MV energy beam.*